\documentclass[nohyper,12pt,letterpaper]{JHEP3}
\usepackage{graphicx,epsfig,youngtab}

\author{Rajsekhar Bhattacharyya$^{1}$, Storm Collins$^{2}$ and Robert de Mello Koch$^{2,3}$\\
\qquad \\
$^{1}$ Department of Physics,\\
Dinabandhu Andrews College,\\
Kolkata-700084,\\
India\\
\qquad \\
$^{2}$ National Institute for Theoretical Physics,\\
Department of Physics and Centre for Theoretical Physics,\\ 
University of the Witwatersrand,\\ 
Wits, 2050,\\ 
South Africa\\
\qquad\\
$^{3}$Stellenbosch Institute for Advanced Studies,\\
Stellenbosch,\\
South Africa\\
\qquad\\
E-mail: \email{rajsekhar@dacollege.org, Storm.Collins@students.wits.ac.za, robert@neo.phys.wits.ac.za}}

\abstract{We argue that restricted Schur polynomials provide a useful parameterization of the complete set of gauge 
invariant variables of multi-matrix models. The two point functions of restricted Schur polynomials are evaluated
exactly in the free field theory limit. They have diagonal two point functions.}

\preprint{WITS-CTP-036}

\title{Exact Multi-Matrix Correlators}


\def \Tr{\mbox{Tr\,}}

\begin{document}

\section{Introduction}

The Maldacena conjecture\cite{Maldacena:1997re}, which claims an equivalence between ${\cal N}=4$ 
super Yang-Mills theory and type IIB string theory in the AdS$_5\times$S$^5$ background, is a significant 
step in establishing the expectation that large $N$ gauge theories are equivalent to string theory. One 
approach towards establishing the conjecture is to implement a direct change of variables from the matrices 
of the gauge theory to the fields of string theory. Collective field theory\cite{Das:1990kaa} provides a 
clear and well defined scheme for making this transition. The first step in this approach is to find a 
useful parameterization of the complete set of gauge invariant variables of the matrix model. For a model 
with more than one matrix, this purely kinematical problem is already nontrivial. In this note, we call
this the {\sl kinematical problem}.

The ${\cal N}=4$ super Yang-Mills theory has six hermittian Higgs fields, $\phi_i$ $i=1,2,...,6$,
transforming in the adjoint of $U(N)$. Form the complex combinations $Z=\phi_1+i\phi_2$, $X=\phi_3+i\phi_4$
and $Y=\phi_5+i\phi_6$. The space of ${1\over 2}$ BPS representations in ${\cal N}=4$ super 
Yang-Mills theory are in one-to-one correspondence with the Schur polynomials built out of $Z$\cite{Corley:2001zk}.
These Schur polynomials have diagonal two point functions\cite{Corley:2001zk}.
Using insights from the dual quantum gravity, excitations of these ${1\over 2}$ BPS states,
{\sl restricted Schur polynomials}, have been identified\cite{Balasubramanian:2004nb}. The restricted Schur polynomial is
obtained by ``attaching'' open string words $W$ to the Schur polynomial. The letters of these open 
string words can be fermions, gauge fields or any of the six Higgs fields. If the word $W$ is to be dual
to an open string, it should contain $O(\sqrt{N})$ letters. If the restricted Schur polynomial contains
$O(N)$ fields, it is dual to a membrane with open strings attached; if it contains $O(N^2)$ fields,
it is dual to a string moving in a new geometry. Thus, the restricted
Schur polynomial starts to address the kinematical problem outlined above. The technology 
for computing correlators of restricted Schur polynomials has enjoyed some 
progress\cite{de Mello Koch:2007uu},\cite{de Mello Koch:2007uv},\cite{Bekker:2007ea}. For related
work see\cite{related}.

In two recent papers, a large class of operators that diagonalize the two point functions in the free 
field theory limit have been given\cite{Kimura:2007wy},\cite{Brown:2007xh}. These include operators built 
from $Z$ and $Z^\dagger$\cite{Kimura:2007wy} and operators built from $X,Y$ and $Z$\cite{Brown:2007xh}. 
Further, the number of such operators matches the number of gauge invariant operators that can be 
constructed. The results of \cite{Kimura:2007wy},\cite{Brown:2007xh} therefore solve the kinematical
problem, in the Higgs sector. This basis also gives a group theoretic way to approach higher point functions
(see \cite{Brown:2007xh} where three and higher point functions are obtained) and to obtain factorization
equations which can be used to build a probability interpretation\cite{Brown:2006zk}. By exploiting
supergroups \cite{Brown:2007xh} have also explained how to include fermions in addition to the Higgs fields.
Finally, the one loop correction to these two points functions has been considered in \cite{Brown:2008rs}.

The purpose of this communication is to argue that the restricted Schur polynomials themselves, provide
a solution to the kinematical problem, in the Higgs sector. This is not unexpected. Indeed,
if one excites a ${1\over 2}$ BPS state by attaching a large number of words that are composed of a single
letter, one is building up multi-matrix operators. Our argument is simple, employing only very basic
group theory. Further, by exploiting the technology already available for restricted Schur polynomials, explicit
formulas for the relevant restricted Schur polynomials and their two point functions are easily obtained.

\section{Two Matrix Model}

Consider a $d=0$ matrix model with two types of complex matrices $A$ and $B$\footnote{The spacetime dependence which has been dropped
from this model can be trivially reinstated using the conformal symmetry of the super Yang-Mills theory.}. These complex matrices act
on an $N$-dimensional vector space $V$, $A:V\to V$. The non-zero correlators are
\begin{equation}
\langle (A)^i_j (A^\dagger)^k_l\rangle =\delta^i_l\delta^k_j =\langle (B)^i_j (B^\dagger)^k_l\rangle .
\label{twopoints}
\end{equation}
Consider the operators
$$ \chi_\alpha = \Tr_{n+m} (O_\alpha A^{\otimes n}\otimes B^{\otimes m}), $$
where $\Tr_{n+m}$ is a trace over $V^{\otimes (n+m)}$. 
$A^{\otimes n}\otimes B^{\otimes m}$ is a shorthand for the tensor 
$A^{i_1}_{j_1}A^{i_2}_{j_2}\cdots A^{i_n}_{j_n}B^{i_{n+1}}_{j_{n+1}}B^{i_{n+2}}_{j_{n+2}}\cdots B^{i_{n+m}}_{j_{n+m}}$
and,
$$\Tr_{n+m} (O_\alpha A^{\otimes n}\otimes B^{\otimes m})=
(O_\alpha)^{j_1 j_2 \cdots j_{n+m}}_{i_1 i_2 \cdots i_{n+m}}
A^{i_1}_{j_1}A^{i_2}_{j_2}\cdots A^{i_n}_{j_n}B^{i_{n+1}}_{j_{n+1}}B^{i_{n+2}}_{j_{n+2}}\cdots B^{i_{n+m}}_{j_{n+m}}.$$
We are interested in computing the correlator $ \langle\chi_\alpha\chi_\beta^\dagger\rangle $. Using (\ref{twopoints}) we obtain
$$ \langle\chi_\alpha\chi_\beta^\dagger\rangle = \sum_{\gamma\in S_n\times S_m}
\Tr_{n+m} (O_\alpha \gamma O_\beta^\dagger \gamma^{-1}).$$
The sum over $\gamma$ is a sum over all possible Wick contractions. 
Assume that
$$O_\beta = \gamma O_\beta\gamma^{-1},\qquad n! m! \, \Tr_{n+m} (O_\alpha O_\beta^\dagger )={\cal N}_\alpha\delta_{\alpha\beta}.$$
This means that the $O_\alpha$ are symmetric branching operators\cite{Kimura:2007wy}.
Then the operators $\chi_\alpha$ diagonalize the two point function
$$ \langle\chi_\alpha\chi_\beta^\dagger\rangle = {\cal N}_\alpha\delta_{\alpha\beta}.$$

We will now argue that a complete set of $O_\alpha$ are given by
$$ O_\alpha = {1\over n! m!}\sum_{\sigma\in S_{n+m}}\Tr_{R_\alpha}(\Gamma_R(\sigma))\sigma ,$$
where $R_\alpha$ is an irreducible representation of $S_n\times S_m$ and $R$ is an irreducible representation of $S_{n+m}$. The $S_n\times S_m$ subgroup
is chosen so that $S_n$ acts on the indices of the $A$s and $S_m$ on the indices of the $B$s. Thus, the $S_n\times S_m$
subgroup that we sum over to include all possible Wick contractions is the same subgroup for which $R_\alpha$ is an
irreducible representation. Under restricting to the
$S_n\times S_m$ subgroup, $R$ will in general be reducible. We can decompose the carrier space of irreducible representation $R$
according to the irreducible $S_n\times S_m$ representations that are subduced. $\Tr_{R_\alpha}$ is an instruction
to trace only over the subspace corresponding to $R_\alpha$. For more details see \cite{de Mello Koch:2007uu}. In this case,
the $\chi_\alpha$ are nothing but restricted Schur polynomials, so that the restricted Schur
polynomials solve the kinematical problem and have diagonal two point functions. 

\noindent
{\it Demonstration that $O_\beta = \gamma O_\beta\gamma^{-1}$:}
\begin{eqnarray} 
\nonumber
\gamma O_\alpha\gamma^{-1} &=&{1\over n! m!}\sum_{\sigma\in S_{n+m}}\Tr_{R_\alpha}(\Gamma_R(\sigma))\gamma\sigma\gamma^{-1}
={1\over n! m!}\sum_{\tau\in S_{n+m}}\Tr_{R_\alpha}(\Gamma_R(\gamma^{-1}\tau\gamma))\tau \\
\nonumber
&=&{1\over n! m!}\sum_{\tau\in S_{n+m}}\Tr_{R_\alpha}(\Gamma_R(\gamma^{-1})\Gamma_R(\tau)\Gamma_R(\gamma))\tau\\
\nonumber
&=&{1\over n! m!}\sum_{\tau\in S_{n+m}}\Tr_{R_\alpha}(\Gamma_{R_\alpha}(\gamma^{-1})\Gamma_R(\tau)\Gamma_{R_\alpha}(\gamma))\tau\\
\nonumber
&=&{1\over n! m!}\sum_{\tau\in S_{n+m}}\Tr_{R_\alpha}(\Gamma_R(\tau))\tau
=O_\alpha .
\end{eqnarray}
We used the fact that $\gamma\in S_n\times S_m$, that $R_\alpha$ is an irreducible representation of $S_n\times S_m$ and that the
trace is invariant under a similarity transformation.

\noindent
{\it Demonstration that $\Tr_{n+m} (O_\alpha O_\beta^\dagger )={\cal N}_\alpha\delta_{\alpha\beta}$:}
\begin{eqnarray}
\nonumber
n! m! \Tr_{n+m} (O_\alpha O_\beta^\dagger )
&=&{1\over n! m!}\sum_{\sigma\in S_{n+m}}\sum_{\tau\in S_{n+m}}\Tr_{R_\alpha}(\Gamma_R(\sigma))
\Tr_{S_\beta}(\Gamma_S(\tau))^* \Tr_{n+m}(\sigma\tau^{-1})\\
\nonumber
&=&{1\over n! m!}\sum_{\sigma\in S_{n+m}}\sum_{\tau\in S_{n+m}}\Tr_{R_\alpha}(\Gamma_R(\sigma))
\Tr_{S_\beta}(\Gamma_S(\tau))^* N^{C(\sigma\tau^{-1})}\\
\nonumber
&=&{1\over n! m!}\sum_{\psi\in S_{n+m}}\sum_{\tau\in S_{n+m}}\Tr_{R_\alpha}(\Gamma_R(\psi\tau))
\Tr_{S_\beta}(\Gamma_S(\tau))^* N^{C(\psi)}.
\end{eqnarray}
Now, lets perform the sum over $\tau$ (use the fact that $(P_{S\to S_\beta})_{jr}^*=(P_{S\to S_\beta})_{rj}$ because the projector
$P_{S\to S_\beta}$ is hermittian)
\begin{eqnarray}
\nonumber
& &\sum_{\tau\in S_{n+m}}\Tr_{R_\alpha}(\Gamma_R(\psi\tau))\Tr_{S_\beta} (\Gamma_S(\tau))^* \\
\nonumber
&=&\sum_{\tau\in S_{n+m}}\sum_{i\,j\,q\,r}(P_{R\to R_\alpha}\Gamma_R(\psi))_{iq}(\Gamma_R(\tau))_{qi}
(P_{S\to S_{\beta}})_{rj}(\Gamma_S(\tau))^*_{rj}\\
\nonumber
&=&\delta_{RS} {(n+m)!\over d_R}\sum_{i\,q}(P_{R\to R_\alpha}\Gamma_R(\psi))_{iq}(P_{S\to S_{\beta}})_{qi}
=\delta_{RS}\delta_{R_\alpha S_\beta}\Tr_{R_\alpha}(\Gamma_R (\psi )){(n+m)!\over d_R} .
\end{eqnarray}
We have used the fundamental orthogonality relation
$$\sum_{\tau\in S_{n+m}}(\Gamma_R(\tau))_{qi}(\Gamma_S(\tau))^*_{rj}={(n+m)!\over d_R}\delta_{qr}\delta_{ij}\delta_{RS}.$$
Thus, using appendix F of \cite{de Mello Koch:2007uu} we obtain
\begin{eqnarray}
\nonumber
n! m! \Tr (O_\alpha O_\beta^\dagger )
&=&{\delta_{RS}\delta_{R_\alpha S_\beta}\over n!m!}{(n+m)!\over d_R}\sum_{\psi\in S_{n+m}}\Tr_{R_\alpha}(\Gamma_R (\psi ))N^{C(\psi )}\\
\nonumber
&=&{\delta_{RS}\delta_{R_\alpha S_\beta}\over n!m!}{(n+m)!\over d_R} d_{R_\alpha}f_R 
=\delta_{RS}\delta_{R_\alpha S_\beta}{{\rm (hooks)}_R\over {\rm (hooks)}_{R_\alpha} }f_R \, .
\end{eqnarray}
$R_\alpha$ is a rep of $S_n\times S_m$ which is labelled by one Young diagram of $n$ boxes, $R_n$, and one Young diagram of
$m$ boxes, $R_m$. ${\rm (hooks)}_{R_\alpha}$ is the product of ${\rm (hooks)}_{R_n}$ with ${\rm (hooks)}_{R_m}$. Arguing as we did 
above, it is simple to obtain
$$ O_\alpha O_\beta ={(n+m)!\over d_R n! m!}\delta_{\alpha\beta} O_\alpha .$$
Thus, up to normalization our operators $O_\alpha$ are projectors. For an earlier use of projectors, along the lines of this note but
in the setting of a single matrix, see \cite{Corley:2002mj}. From now on
we write $\chi_{R,R_\alpha}$ instead of $\chi_\alpha$. In general, the row and column index of the restriction $R_\alpha$ can be different
(see \cite{Balasubramanian:2004nb},\cite{de Mello Koch:2007uu} for a detailed discussion). Spell out these row and column indices by
replacing $R_\alpha\to (r_{\alpha_1},r_{\alpha_2})$. The two point function is
\begin{equation}
\langle \chi_{R,(r_{\alpha_1},r_{\alpha_2})}\chi_{S,(s_{\beta_1},s_{\beta_2})}^\dagger\rangle=\delta_{RS}
\delta_{r_{\alpha_1}s_{\beta_1}}\delta_{r_{\alpha_2}s_{\beta_2}}{{\rm (hooks)}_R\over {\rm (hooks)}_{R_\alpha} }f_R\, .
\label{result}
\end{equation}
It is equally easy to argue that
\begin{equation}
\langle \chi_{R,(r_{\alpha_1},r_{\alpha_2})}\chi_{S,(s_{\beta_1},s_{\beta_2})}\rangle=\delta_{RS}
\delta_{r_{\alpha_1}s_{\beta_2}}\delta_{r_{\alpha_2}s_{\beta_1}}{{\rm (hooks)}_R\over {\rm (hooks)}_{R_\alpha} }f_R\, .
\label{secondresult}
\end{equation}
In sections 3 and 5 we will give evidence that the number of restricted Schur polynomials $\chi_{R,R_\alpha}$ is equal to the
number of gauge invariant operators in the matrix model.

\section{Counting}

The number of gauge invariant operators $N(n,m)$ built out of $n$ $A$s and $m$ $B$s is given by Polya theory as
$$\prod_{k=1}^\infty {1\over 1-(x^k+y^k)}=\sum_{n,m} N(n,m) x^n y^m .$$
We claim that the number of gauge invariant operators $N(n,m)$ is equal to the number of restricted Schur polynomials
$\chi_{R,R_\alpha}$ with $R$ an irreducible representation of $S_{n+m}$ and $R_\alpha$ an irreducible representation of $S_n\times S_m$. It is easy to check for
small values of $n$ and $m$ that this is indeed the case. As an example, consider $m=n=2$. In this case, $R$ is an irreducible representation
of $S_4$. We easily find $N(2,2)=10$. The allowed restricted traces $(R;R_\alpha)$ are
$$ (\yng(4);\yng(2)\otimes\yng(2))$$
$$ (\yng(3,1);\yng(2)\otimes\yng(2)),\quad (\yng(3,1);\yng(2)\otimes\yng(1,1)),\quad (\yng(3,1);\yng(1,1)\otimes\yng(2))$$
$$(\yng(2,2);\yng(2)\otimes\yng(2)),\quad (\yng(2,2);\yng(1,1)\otimes\yng(1,1))$$
$$(\yng(2,1,1);\yng(1,1)\otimes\yng(2)),\quad (\yng(2,1,1);\yng(1,1)\otimes\yng(1,1))(\yng(2,1,1);\yng(2)\otimes\yng(1,1))$$
$$(\yng(1,1,1,1);\yng(1,1)\otimes\yng(1,1)).$$
Thus, there are indeed ten possible restricted Schur polynomials.

There is a subtlety that did not show up in the above example: in the notation of 
\cite{de Mello Koch:2007uu},\cite{Balasubramanian:2004nb}, we can trace over an off the diagonal block.
For example, amoung the $S_3\times S_3$ irreducible representations subduced by the $S_6$ irreducible representation
$$ R=\yng(3,2,1)$$
we find two copies of
$$ \yng(2,1)\otimes\yng(2,1) .$$
Call these two copies $R_\alpha^{(1)}$ and $R_\alpha^{(2)}$. When performing the restricted trace, we can use $R_\alpha^{(i)}$
for the row index and $R_\alpha^{(j)}$ for the column index with $R_\alpha^{(i)}\ne R_\alpha^{(j)}$. Thus, there are four
possible operators we can define. In general, if $R$ subduced $m$ copies of an irreducible representation $R_\alpha$ we would be able to
construct $m^2$ independent operators. For further details consult section 2.2 of \cite{de Mello Koch:2007uu}.

\section{Examples}

The simplest way to construct restricted Schur polynomials, is to use a projection operator to implement the restricted
trace. In this section we will construct restricted Schur polynomials built from at most three matrices, which can be any
of two different types $X$ or $Y$. This will already allow us to see that the restricted Schur polynomials define a different 
basis for gauge invariant operators, than the bases given in \cite{Kimura:2007wy},\cite{Brown:2007xh}. The construction of
$$\chi_{\tiny \yng(2);\yng(1)^\otimes\yng(1)}=\Tr (X) \Tr (Y)   + \Tr (XY),\qquad 
  \chi_{\tiny \yng(1,1);\yng(1)^\otimes\yng(1)}=\Tr (X) \Tr (Y) - \Tr (XY),$$
is particularly simple because we do not need a projector to implement the restricted trace. This follows because $\yng(1)^\otimes\yng(1)$
is the only $S_1\times S_1$ irreducible representation subduced from either $\yng(2)$ or $\yng(1,1)$.
Up to normalization, these are identical to the operators constructed in appendix E1 of \cite{Brown:2007xh}.
Consider next
$$\chi_{\tiny \yng(3);\yng(2)^\otimes\yng(1)}={1\over 2}\left[ \Tr (X)^2 \Tr (Y)+\Tr (X^2)\Tr (Y)+2\Tr (XY)\Tr (X)+2\Tr (X^2Y)\right],$$
$$\chi_{\tiny \yng(1,1,1);\yng(1,1)^\otimes\yng(1)}={1\over 2}\left[ \Tr (X)^2 \Tr (Y)-\Tr (X^2)\Tr (Y)-2\Tr (XY)\Tr (X)+2\Tr (X^2Y)\right].$$
For these two restricted Schur polynomials we again do not need a projector to implement the restricted trace. If we take
$$\chi_{R,R_\alpha}={1\over 2!}\sum_{\sigma\in S_3}\Tr_{R_\alpha}(\Gamma_R(\sigma ))X^{i_1}_{i_{\sigma(1)}}
X^{i_2}_{i_{\sigma(2)}} Y^{i_3}_{i_{\sigma(3)}},$$
then $R_\alpha$ is an irreducible representation of $S_2\times S_1$. The $S_2$ subgroup is obtained by taking those elements
of $S_3$ that act on the indices of the $X$s, i.e. $\{ 1,(12)\}$. To compute
$$\chi_{\tiny \yng(2,1);\yng(2){}^\otimes\yng(1)}= {1\over 2}\left[ \Tr (X)^2 \Tr (Y)+\Tr (X^2)\Tr (Y)-\Tr (XY)\Tr (X)-\Tr (X^2Y)\right],$$
we used the projector
$$ P_{\tiny \yng(2,1)\to \yng(2){}^\otimes\yng(1)}={1\over 2}\left(1+\Gamma_{\tiny \yng(2,1)}\big( (12)\big)\right).$$
To compute
$$\chi_{\tiny \yng(2,1);\yng(1,1){}^\otimes\yng(1)}= {1\over 2}\left[ \Tr (X)^2 \Tr (Y)-\Tr (X^2)\Tr (Y)+\Tr (XY)\Tr (X)-\Tr (X^2Y)\right],$$
we used the projector
$$ P_{\tiny \yng(2,1)\to \yng(1,1){}^\otimes\yng(1)}={1\over 2}\left(1-\Gamma_{\tiny \yng(2,1)}\big( (12)\big)\right).$$
For more details on these projectors see appendix B.2 of \cite{de Mello Koch:2007uu} and appendix A of \cite{Bekker:2007ea}.
Comparing these expressions to the expressions in appendix E.2 of \cite{Brown:2007xh}, it is clear that the basis furnished
by the restricted Schur polynomials does not coincide with the basis of \cite{Brown:2007xh}.

We can use the $\Sigma$ map of \cite{Kimura:2007wy} to construct new operators built out of $Z$ and $Z^*$. Under the map $\Sigma$, 
$B_\alpha=\Sigma^{-1}(O_\alpha)$ becomes a sum over elements of the Brauer algebra. In \cite{Kimura:2007wy} it was argued that if 
$\gamma O_\alpha\gamma^{-1}=O_\alpha$ for $\gamma\in S_n\times S_m$ then $\gamma B_\alpha\gamma^{-1}=B_\alpha$ for $\gamma\in S_n\times S_m$. 
Also, again using a result of \cite{Kimura:2007wy}, ($\Tr_{m+n}$ denotes a trace over $V^{\otimes (n+m)}$ and $\Tr_{m,n}$ denotes a trace over 
$V^{\otimes n}\otimes\bar{V}^{\otimes m}$)
$$\Tr_{m,n}(B_\alpha B_\beta)=\Tr_{m+n}(O_\alpha O_\beta)={{\cal N}_\alpha\over n!m!}\delta_{\alpha\beta}.$$
Thus, the operators 
$$ \eta_\alpha=\Tr_{m,n}(B_\alpha Z^{\otimes n}\otimes Z^{*\,\otimes m}),$$
have a diagonal two point function
$$\langle \eta_\alpha \eta_\beta^\dagger\rangle = {\cal N}_\alpha\delta_{\alpha\beta}.$$
For $m=n=1$ we find
$$ B_1=\Sigma ({1\over 2}\Big(1+(12))\Big)={1\over 2}\Big( 1+C_{1\bar{1}}\Big),$$
$$ B_2=\Sigma ({1\over 2}\Big(1-(12))\Big)={1\over 2}\Big( 1-C_{1\bar{1}}\Big).$$
These do not match the operators given in appendix A.4.1 of \cite{Kimura:2007wy}, implying that the restricted Schur polynomials 
do not coincide with the basis constructed in \cite{Kimura:2007wy} either. This is clear when one notes that the coefficients on 
the projectors in \cite{Kimura:2007wy} are $N$ dependent; there is no way in which our operators could pick up $N$ dependent 
coefficients.

Recall that weights are assigned to boxes in a Young diagram by assigning $N$ to the box in the upper left hand corner of
the Young diagram, adding one each time we move to the right and subtracting one each time we move down. Thus, box $i$ in
the Young diagram
$$\young(123,45)$$
has weight $c_i$ with $c_1=c_5=N$, $c_2=N+1$, $c_3=N+2$ and $c_4=N-1$. $f_R$ is the product of weights of the Young diagram,
so that, for example
$$ f_{\yng(3,2)}=N^2 (N^2-1)(N+2). $$
Next, since
$$ {\rm hooks}(\yng(5))=5!\quad {\rm hooks}(\yng(3))=3!\quad {\rm hooks}(\yng(2))=2!,$$
we have from (\ref{result})
$$\langle\chi_{\tiny \yng(5);\yng(3){}^{\otimes}\yng(2)}\chi_{\tiny \yng(5);\yng(3){}^{\otimes}\yng(2)}^\dagger\rangle
={5!\over 3!\times 2!}f_{\tiny\yng(5)}.$$
Similary,
$$\langle\chi_{\tiny \yng(3,3);\yng(3){}^{\otimes}\yng(3)}\chi_{\tiny \yng(3,3);\yng(3){}^{\otimes}\yng(3)}^\dagger\rangle
={4!\times 3!\over 3!\times 3!}f_{\tiny\yng(3,3)}.$$
If any of the labels on the restricted Schur polynomial do not match, the correlator vanishes
$$\langle\chi_{\tiny \yng(4,1);\yng(3){}^{\otimes}\yng(2)}\chi_{\tiny \yng(5);\yng(3){}^{\otimes}\yng(2)}^\dagger\rangle
=0,$$
$$\langle\chi_{\tiny \yng(4,1);\yng(3){}^{\otimes}\yng(2)}\chi_{\tiny \yng(4,1);\yng(3){}^{\otimes}\yng(1,1)}^\dagger\rangle
=0,$$
$$\langle\chi_{\tiny \yng(3,3);\yng(3){}^{\otimes}\yng(3)}\chi_{\tiny \yng(3,3);\yng(2,1){}^{\otimes}\yng(3)}^\dagger\rangle
=0.$$

To determine which $S_n\times S_m$ irreducible representations are subduced by a particular $S_{n+m}$ irreducibe representation
is easy: assume that the Young diagram $R$ describes the irreducible representation of $S_{n+m}$ that we are studying. Consider
all possible ways of removing $n$ boxes from $R$ so that the remaining $m$ boxes form a legal Young diagram $R_m$. Remove the
$n$ boxes preserving common sides and take the tensor product of the removed pieces to get $R_n$. This rule is easily
illustrated with an example; consider 
$$ R=\yng(3,2,1). $$
Assume that $n=m=3$. Denoting removed boxes with an $x$ we have
\begin{eqnarray}
\nonumber
&&\young({\,}{\,}{\,},{x}{x},{x})\qquad R_m=\yng(3),\quad R_n=\yng(2,1),\\  
\nonumber
&&\young({\,}{x}{x},{\,}{x},{\,})\qquad R_m=\yng(1,1,1),\quad R_n=\yng(2,1),\\
\nonumber
&&\young({\,}{\,}{x},{\,}{x},{x})\qquad R_m=\yng(2,1),\qquad R_n=\yng(1)\otimes\yng(1)\otimes\yng(1)=\yng(3)\oplus\yng(1,1,1)
\oplus\, 2\, \yng(2,1).
\end{eqnarray}
Thus, $R$ subduces 6 irreducible representations of $S_3\times S_3$.

\section{Generalization to Multi-Matrix Models}

The above results generalize in a simple way to multi-matrix models. Consider a model of $M$ matrices and assume that 
$\chi_{R,R_\alpha}$ is built from $m_i$ matrices of each type. Then $R_\alpha$ is an irreducible representation of 
$S_{m_1}\times S_{m_2}\times\cdots\times S_{m_M}$. To remove self contractions (present if we have real matrices or
if we build $\chi_{R,R_\alpha}$ from complex matrices and their adjoints) we simply normal order $\chi_{R,R_\alpha}$.
This gives a unified treatment of both branes/antibrane systems and operators built from more than one Higgs field. 
These operators are labeled by $M+1$ Young diagrams, one with $m_1+m_2+ ... +m_M$ boxes, $R$ and $M$ with $m_i$ boxes, $R_i$.
In this more general case we still have (\ref{result}) with
$$ {\rm (hooks)}_{R_\alpha} = \prod_{i=1}^M {\rm (hooks)}_{R_i}.$$
It is straight forward to replace boxes in the $R_i$ by open strings so that excited operators can be constructed and
studied using the techniques developed in \cite{de Mello Koch:2007uu},\cite{de Mello Koch:2007uv},\cite{Bekker:2007ea}.

We again claim that the total number of restricted Schur polynomials that can be 
defined will be equal to the number of gauge invariant operators that can be constructed. There are some non-trivial
tests we can perform of this claim. For example, consider operators built using one of each of the $M$ types. In this case, 
we need to start with an irreducible representation of $S_M$ and count how many restricted Schur polynomials we can
form when the representation of the restriction is $S_1\times S_1\times \cdots \times S_1$ (there are $M$ factors).
To get the number of irreducible representations that can be subduced from a given Young diagram $R$, we need to count the
number of ways we can pull boxes off $R$ such that at each step we have a legal Young diagram. This is obviously $d_R$,
the dimension of the $S_M$ representation labeled by $R$. Any of these subduced representations may be twisted, so that
we obtain a total of $d_R^2$ operators. Thus, the total number of restricted Schur polynomials, found by summing over all 
$S_M$ irreducible representations, is simply
$$\sum_R (d_R)^2 = M! .$$
Lets now compare this to the counting of the gauge invariant operators. According to Polya theory, the number of
gauge invariant operators is given by the coefficient of $x_1 x_2 \,\cdots \,x_M$ in the expansion of
$$\prod_{k=1}^\infty {1\over 1-(x_1^k+x_2^k+\cdots +x_M^k)}=\sum_{n_1,n_2,\cdots,n_m}
t(n_1,n_2,\cdots ,n_M) x_1^{n_1}x_2^{n_2}\cdots x_M^{n_M}.$$
It is simple to see that
$$ t(1,1,\cdots ,1)=M!,$$
which supports our claim.

\section{Numerical Tests}

We have counted the number of restricted Schur polynomials $\chi_{R,R_\alpha}$ that be obtained when $R$ is an
irreducible representation of $S_n$ with $n\le 6$ and we have a total of $M=6$ matrices. In all of these cases,
the number of restricted Schur polynomials equals the number of gauge invariant operators counted using Polya
theory. Further, we have numerically evaluated the two point functions of these restricted Schur polynomials
and verified that (\ref{result}) is indeed correct. In performing these checks, the restricted characters 
$\Tr_{R_\alpha}\left(\Gamma_{R}\big[\sigma\big]\right)$ were evaluated by explicitly constructing the matrices 
$\Gamma_{R}\big[\sigma\big]$. Each representation used was obtained by induction. One induces a reducible 
representation; the irreducible representation required was isolated using projection operators built from the 
Casimir obtained by summing over all two cycles. The restricted trace was then evaluated with the help of suitable 
projectors. See appendix B.2 of \cite{de Mello Koch:2007uu} and appendix A of \cite{Bekker:2007ea} for more details. 
In all cases the numerical result is in exact agreement with (\ref{result}).

\section{Conclusions}

Restricted Schur polynomials provide a useful parameterization of the complete set of gauge 
invariant variables of multi-matrix models. They have diagonal two point functions. Since in
the labeling of the restricted Schur polynomial, each type of matrix has its own Young diagram, 
the technology for attaching open strings has a straight forward generalization to the operators
considered in this article.

For brane-anti-brane operators, the restricted Schur polynomials do not coincide with the Brauer basis 
constructed in \cite{Kimura:2007wy}. Since the Brauer projectors are $N$ dependent, the relation between 
the two bases is $N$ dependent. It seems that the Brauer basis may be the most useful for identifying
brane - anti-brane operators and the restricted Schur polynomial basis for stringy excitations. It is 
plausible that there is a simple relation between the restricted Schur polynomials and the operators of
\cite{Brown:2007xh}. For example, 
$\chi_{\tiny \yng(2,1);\yng(2){}^\otimes\yng(1)}-\chi_{\tiny \yng(2,1);\yng(1,1){}^\otimes\yng(1)}$
is (up to an overall constant factor) equal to the operators constructed in E.2 and E.3 of \cite{Brown:2007xh}.
Since the restricted Schur polynomials have an interpretation in terms of attaching open strings, developing
this relation may well shed light on the interpretation of the labels of the operators constructed in \cite{Brown:2007xh}.
We leave this interesting problem for the future.

Finally, it would be interesting to explore finite $N$ effects. These effects are encoded in the fact that 
our Young diagram labels can have at most $N$ rows. Specifically, in the restricted Schur polynomial
$\chi_{R,(r_{\alpha_1},r_{\alpha_2})}$ we must require that the Young diagram $R$ has at most $N$ rows;
the same will automatically be true for $r_{\alpha_1}$ and $r_{\alpha_2}$. This should translate into a 
generalization of the stringy exclusion principle present for Schur polynomials built using a single matrix $Z$. 
Finite $N$ counting for multi-matrix operators has been considered in \cite{Brown:2007xh},\cite{Dolan:2007rq}. 
For example, the number of operators built using $\mu_1$ $X$ fields and $\mu_2$ $Y$ fields, at infinite
$N$ is given by\footnote{We are considering the case of two matrices for simplicity. The formula for $M$ matrices
has been determined in \cite{Brown:2007xh},\cite{Dolan:2007rq}.}
$$ N(\mu_1,\mu_2)=\sum_T\sum_\Lambda C(T,T,\Lambda)g(\mu ;\Lambda). $$
In this formula, $T$ is a representation of $S_n$ with $n=\mu_1+\mu_2$, $C(T,T,\Lambda )$ is the coefficient of $\Lambda$ in
the (inner) tensor product $T\otimes T$ and $g(\mu ;\Lambda)$ is the Littlewood-Richardson coefficient which counts states
in the representation $\Lambda$ that have the field content $\mu = \big[\mu_1\big]\otimes\big[\mu_2\big].$ To get the finite
$N$ counting, one simply truncates the sum over $T$ to Young diagrams with at most $N$ rows.
We can see, in some simple examples, that our cut off on $R$ does indeed match the finite $N$ counting of 
\cite{Brown:2007xh},\cite{Dolan:2007rq}.
Consider for example the operators built using 3 $X$ fields and a single $Y$ field. The relevant Littlewood-Richardson coefficients
are
$$ g({\tiny \yng(3),\yng(1);\yng(4)})=1,\qquad g({\tiny \yng(3),\yng(1);\yng(3,1)})=1 .$$
The relevant inner products are
{\tiny
$$\yng(4)\otimes \yng(4) = \yng(4),$$
$$ \yng(3,1)\otimes\yng(3,1) = \yng(4) \oplus \yng(3,1) \oplus \yng(2,2) \oplus \yng(2,1,1),$$
$$ \yng(2,2)\otimes\yng(2,2) = \yng(4) \oplus \yng(2,2) \oplus \yng(1,1,1,1),$$
$$ \yng(2,1,1)\otimes\yng(2,1,1)=\yng(4) \oplus \yng(3,1) \oplus \yng(2,2) \oplus \yng(2,1,1),$$
$$\yng(1,1,1,1)\otimes\yng(1,1,1,1) = \yng(4).$$
}
Clearly then, at infinite $N$, the number of operators we can construct is
$$ N(2,1)=\sum_T (C(T,T,{\tiny \yng(4)})+C(T,T,{\tiny \yng(3,1)}))$$
$$=C({\tiny \yng(4),\yng(4),\yng(4)})+C({\tiny \yng(3,1),\yng(3,1),\yng(4)})+C({\tiny \yng(2,2),\yng(2,2),\yng(4)})
+C({\tiny \yng(2,1,1),\yng(2,1,1),\yng(4)})$$
$$+C({\tiny \yng(1,1,1),\yng(1,1,1),\yng(4)})
+C({\tiny \yng(3,1),\yng(3,1),\yng(3,1)})+C({\tiny \yng(2,1,1),\yng(2,1,1),\yng(3,1)})=7.$$
At $N=2$, this counting becomes
$$ N(2,1)=\sum_T (C(T,T,{\tiny \yng(4)})+C(T,T,{\tiny \yng(3,1)}))$$
$$=C({\tiny \yng(4),\yng(4),\yng(4)})+C({\tiny \yng(3,1),\yng(3,1),\yng(4)})+C({\tiny \yng(2,2),\yng(2,2),\yng(4)})$$
$$+C({\tiny \yng(3,1),\yng(3,1),\yng(3,1)})=4.$$
Lets now count the restricted Schur polynomials. At infinite $N$ we find 7 possible operator, with $R,r_{\alpha_1}r_{\alpha_2}$ labels
given by
{\tiny
$$\yng(4),\yng(3)\,\yng(1)\qquad \yng(3,1),\yng(3)\,\yng(1)\qquad \yng(3,1),\yng(2,1)\,\yng(1)\qquad \yng(2,2),\yng(2,1)\,\yng(1) $$
$$\yng(2,1,1),\yng(2,1)\,\yng(1)\qquad \yng(2,1,1),\yng(1,1,1)\,\yng(1)\qquad \yng(1,1,1,1),\yng(1,1,1)\,\yng(1) $$
}
At $N=2$ there are only 4 operators, with labels given by
{\tiny
$$\yng(4),\yng(3)\,\yng(1)\qquad \yng(3,1),\yng(3)\,\yng(1)\qquad \yng(3,1),\yng(2,1)\,\yng(1)\qquad \yng(2,2),\yng(2,1)\,\yng(1) $$
}
Providing a proof that our cut off on $R$ matches the finite $N$ counting of \cite{Brown:2007xh},\cite{Dolan:2007rq} remains
an interesting open problem.

{\vskip 1.0cm}

\noindent
{\it Acknowledgements:} We would like to thank Norman Ives, Sanjaye Ramgoolam and Alex Welte for enjoyable, 
helpful discussions and/or correspondence. We would also like to thank Sanjaye Ramgoolam for a careful
reading of the manuscript and Tom Brown for useful correspondence on the first version  of this article. 
Finally, we would like to thank Norman Ives for again pointing out which is our left hand and which is 
our right hand. This work is based upon research supported by the South African Research Chairs
Initiative of the Department of Science and Technology and National Research Foundation.
Any opinion, findings and conclusions or recommendations expressed in this material
are those of the authors and therefore the NRF and DST do not accept any liability
with regard thereto. This work is also supported by NRF grant number Gun 2047219.

\end{document}